\documentclass[preprint,amsmath]{revtex4}
\usepackage{psfrag}
\usepackage[dvips]{graphicx}

\begin{document}

\title{Streamwise oscillation of spanwise velocity \\
at the wall of a channel for turbulent drag reduction}
\author{Claudio Viotti$^1$}
\author{Maurizio Quadrio$^1$}
\affiliation{$^1$Dipartimento di Ingegneria Aerospaziale del Politecnico di Milano
\\ via La Masa 34 - 20156 Milano, Italy}
\author{Paolo Luchini$^2$}
\affiliation{$^2$Dipartimento di Ingegneria Meccanica Universit\`a di Salerno
\\ 84084 Fisciano (SA), Italy}
\date{\today}

\newcommand{\pd}[2]{\frac{\partial #1 }{\partial #2}}
\newcommand{\ppd}[2]{\frac{\partial^2 #1}{{\partial #2}^2}}
\newcommand{\ty}{\tilde{y}}
\newcommand{\Ai}{Ai}
\newcommand{\Bi}{Bi}
\newcommand{\ud}{\mathrm{d}}
\newcommand{\sP}{\mathcal{P}}
\newcommand{\sU}{\mathcal{U}}
\newcommand{\aver}[1]{\left\langle #1 \right\rangle}
\newcommand{\rme}{\mathrm{e}}

\begin{abstract}
Steady forcing at the wall of a channel flow is studied via DNS to assess its ability of yielding reductions of turbulent friction drag. The wall forcing consists of a stationary distribution of spanwise velocity that alternates in the streamwise direction. The idea behind the forcing builds upon the existing technique of the spanwise wall oscillation, and exploits the convective nature of the flow to achieve an unsteady interaction with turbulence. 

The analysis takes advantage of the equivalent laminar flow, that is solved analytically to show that the energetic cost of the forcing is unaffected by turbulence. In a turbulent flow, the alternate forcing is found to behave similarly to the oscillating wall; in particular an optimal wavelength is found that yields a maximal reduction of turbulent drag. The energetic performance is significantly improved, with more than 50\% of maximum friction saving at large intensities of the forcing, and a net energetic saving of 23\% for smaller intensities.

Such a steady, wall-based forcing may pave the way to passively interacting with the turbulent flow to achieve drag reduction through a suitable distribution of roughness, designed to excite a selected streamwise wavelength.
\end{abstract}

\maketitle

\section{Introduction}

In recent years several attempts at controlling turbulence through a number of wall-based forcing methods have been reported \cite{gadelhak-2000}, often aimed at frictional drag reduction (DR). A large number of works, exploiting both numerical and experimental approaches, has been devoted to this goal.

In this paper we focus on spanwise wall-based forcing, i.e. on a class of forcing methods designed to modify favorably the turbulent flow by introducing an external action directed in the spanwise direction. Early work addressing the modification of wall turbulence by creating a cross flow can be traced back to Bradshaw \& Pontikos \cite{bradshaw-pontikos-1985}. Spanwise forcing of turbulent flows has been reviewed by Karniadakis \& Choi \cite{karniadakis-choi-2003}. In 1992, Jung, Mangiavacchi \& Akhavan \cite{jung-mangiavacchi-akhavan-1992} introduced the spanwise-oscillating wall technique, where the wall of a fully turbulent channel flow is subject to an alternate harmonic motion in the spanwise direction. If $W$ indicates the spanwise velocity component at the wall, the law that defines this forcing method is:
\begin{equation}
\label{eq:time}
  W = A \sin \left( \frac{2\pi}{T} t \right)
\end{equation}
where $t$ is time, and $A$ and $T$ are the oscillation amplitude and period, respectively. Once $A$ and $T$ are set within the optimum range from the viewpoint of DR performance, the authors observed, by means of direct numerical simulations (DNS), that a strong suppression of turbulence occurs in the wall region, accompanied by a significant reduction of the mean friction.

The analysis has been carried on in successive studies: we recall Quadrio \& Sibilla \cite{quadrio-sibilla-2000} for the pipe flow, and J.-I. Choi, Xu \& Sung  \cite{choi-xu-sung-2002} and Quadrio \& Ricco \cite{quadrio-ricco-2003,quadrio-ricco-2004}, as DNS-based analyses which contributed new datasets and detailed descriptions of the flow. Laboratory experiments, due to Laadhari {\it et al.} \cite{laadhari-skandaji-morel-1994} and K. S. Choi \cite{choi-graham-1998,choi-2002}, complemented the numerical works and addressed the issue of dependency of DR on the value of the Reynolds number. Alltogether these works have contributed to demonstrating that the natural friction drag of the turbulent flow can be reduced (at least at moderate values of $Re$) up to 45\% for $A^+ \approx 25$ (quantities with the $^+$ superscript are made dimensionless with viscous wall units). An optimal value of the oscillation period exists, namely $T_{opt}^+ = 100 - 125$, that yields the maximum DR at all amplitudes. The net energy saving, that substracts from the reduced flow-driving pumping power the power expense required to move the wall against viscous resistance, has been addressed first by Baron and Quadrio \cite{baron-quadrio-1996}, and today it is recognized \cite{karniadakis-choi-2003} that it can reach up to 10\%. Crediting a commonly accepted qualitative explanation, the transverse oscillating boundary layer induced by the wall motion explains the drag reduction, since it produces a phase displacement between the wall-layer turbulence structures, capable of weakening the viscous wall cycle. When phase-averaged, this time-alternating layer in the turbulent regime has been found to coincide with the oscillating laminar Stokes layer, for which an analytic solution is known as a classic solution of the boundary layer equations \cite{schlichting-gersten-2000}. This has been recently exploited to determine a parameter \cite{quadrio-ricco-2004} that is capable of scaling linearly with DR and thus makes it possible to predict DR capabilities.

Advantages and drawbacks of the oscillating-wall technique are obvious. It presents energetic performance that could make it worth of practical implementations, and thanks to its open-loop character it does not need distributed sensors or actuators, that would be still unpractical with the technology available today. On the other hand, by its very nature this technique requires moving parts, and thus does not lend itself to be implemented as a passive device, which is on the other hand the most appealing possibility application-wise.

In this paper, we aim at extending the oscillating-wall technique, to take one further step towards the long-term objective of a successful practical implementation; in particular we want to translate the time-dependent forcing law expressed by Eq.(\ref{eq:time}) into a stationary formula. This goal can be achieved by exploiting the convective nature of wall-bounded flows. Though at the wall the mean velocity profile annihilates, it is well known that the convection velocity of a turbulent wall flow (or, more precisely, the convection velocity of turbulent fluctuations) resembles the mean velocity profile only in the bulk of the flow. Kim \& Hussain \cite{kim-hussain-1993} have shown some years ago that near the wall, say below $y^+ = 15$, the convection velocity becomes essentially independent upon the wall distance, and remains constant at the value $\sU_w^+ \approx 10$. It is this near-wall value of the convection velocity that will enable us to translate the temporal forcing from Eq.(\ref{eq:time}) into a spatially-oscillating forcing, yielding the following forcing law that is expected to modify the turbulent flow in a similar manner to the oscillating wall:

\begin{equation}
\label{eq:space}
W = A \sin \left( \frac{2\pi}{\lambda_x} x \right).
\end{equation}

\begin{figure}
\centering
\includegraphics[width=.7\columnwidth]{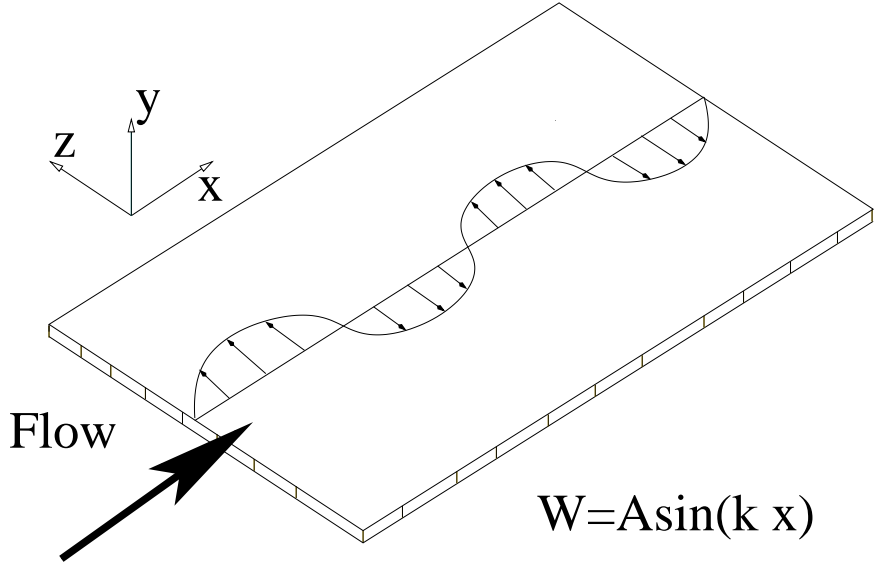}
\caption{Sketch of the wall spanwise forcing discussed in the present work.}
\label{fig:sketch}
\end{figure}
Here $x$ denotes the streamwise coordinate, and $\lambda_x$ is the forcing wavelength. 
The resulting distribution of spanwise wall velocity is sketched in fig.\ref{fig:sketch}, where the coordinate system employed in this paper is also indicated.

Though the control law (\ref{eq:space}) has never been considered in the literature, one paper where a similar space-time extension has been discussed in the past is the one by Berger and coworkers \cite{berger-etal-2000}. In a parametric DNS-based study a spanwise-oriented Lorentz volume force was simulated to obtain the following two forcing configurations:
\begin{equation}
\label{eq:berger-time}
F_z = B \rme^{-y/\Delta} \sin \left( \frac{2\pi}{T} t \right),
\end{equation}
\begin{equation}
\label{eq:berger-space}
F_z = B \rme^{-y/\Delta}\sin \left( \frac{2\pi}{\lambda_x} x \right),
\end{equation}
where $\Delta$ is the penetration depth of the forcing, and $B$ its intensity. For certain values of the parameters, the streamwise-dependent law (\ref{eq:berger-space}) produced a DR comparable to that of the time-dependent law (\ref{eq:berger-time}), with the added benefit of an improved energetic balance. However, the difference between a body force and a wall-based forcing is substantial, as discussed for example by Zhao et al. \cite{zhao-wu-luo-2004}, and no definite conclusion can thus be drawn {\em a priori} from the study by Berger et al. \cite{berger-etal-2000} with respect to the oscillating velocity, which is strictly a wall-based forcing. Indeed, it will be shown later that some of the results described in Ref. \cite{berger-etal-2000} (for example the existence of an optimal wavelength that depends on the forcing intensity) do not apply at all to the type of forcing considered here.
Schoppa \& Hussain \cite{schoppa-hussain-1998} showed that a significant amount of drag reduction can be achieved by introducing in a channel flow a spanwise velocity gradient, generated in the DNS by a large-scale, streamwise-aligned and $x$-independent rolls. Such velocity gradient weakens the near-wall cycle by suppressing the transient growth of streaks that would otherwise be stable according to normal-mode analysis. This mechanism is then addressed in a subsequent paper \cite{schoppa-hussain-2002}, where the role of streamwise-varying spanwise perturbations is highlighted in the context of the turbulence regeneration cycle. The dominant wavelength of the streak waviness is found to be very similar to the streamwise length of the coherent structures educed from conditional analysis of turbulent flow fields \cite{jeong-etal-1997}, namely about 300--400 viscous units.

Aim of the present paper is to investigate the DR and energetic performance of the steady control law (\ref{eq:space}). Thanks to an accurate data set, purposely obtained by several DNS simulations, the $\lambda_x-A$ parameter space will be explored in detail. The value of the Reynolds number will be fixed at $Re_\tau=200$ (based on the friction velocity of the reference flow and half the channel gap); any dependence on this flow parameter is not discussed here. The effects of the forcing (\ref{eq:space}) on a laminar channel flow will be preliminarly studied. We will see that a laminar solution, although approximated, can be obtained in analytical form, and this solution will then be used to help understanding the DR properties of the forcing (\ref{eq:space}) when used on a turbulent flow. In the end, we will be able to connect this forcing to a physical application that does not necessarily involve a moving wall.

The paper is organized as follows. Sec.\ref{sec:laminar} contains a theoretical discussion of the associated laminar flow, that is useful for predicting some of the energetic characteristics of the present technique (further analytical details are deferred to an Appendix). In Sec.\ref{sec:num} the numerical simulations of the turbulent case and their discretization parameters are described. Results about DR and energetic performances are reported and discussed in Sec.\ref{sec:results}, together with a visual and statistical description of the forced flow, compared to the reference unperturbed one. Some conclusive remarks, including examples of how the present forcing can be implemented in practice, are given in the concluding Sec.\ref{sec:conclusions}.

\section{Laminar flow}
\label{sec:laminar}

We consider first the laminar flow in a plane channel subject to either the temporal boundary forcing (\ref{eq:time}) or the spatial boundary forcing (\ref{eq:space}). For the temporal case, it is easy to show that the incompressible momentum equation for the spanwise component $w$ decouples, so that the streamwise flow is described by the classic Poiseuille parabolic solution, and the entire flow consists in this parabolic profile plus a spanwise alternating motion, that is identical to the oscillating transversal boundary layer that develops in a still fluid bounded by a wall subject to harmonic oscillation (i.e. the so-called Stokes second problem, that possesses a classic analytical solution \cite{schlichting-gersten-2000}). This oscillating boundary layer will be called Temporal Stokes Layer (TSL) in the following.

We address now the laminar flow subject to the space-varying boundary forcing (\ref{eq:space}), to define and discuss the spatial equivalent of the TSL, that will be called Spatial Stokes Layer (SSL). We leave some analytical details to Appendix \ref{sec:analytical}, but the main ideas are given here as follows.

\begin{figure}
\includegraphics[width=\columnwidth]{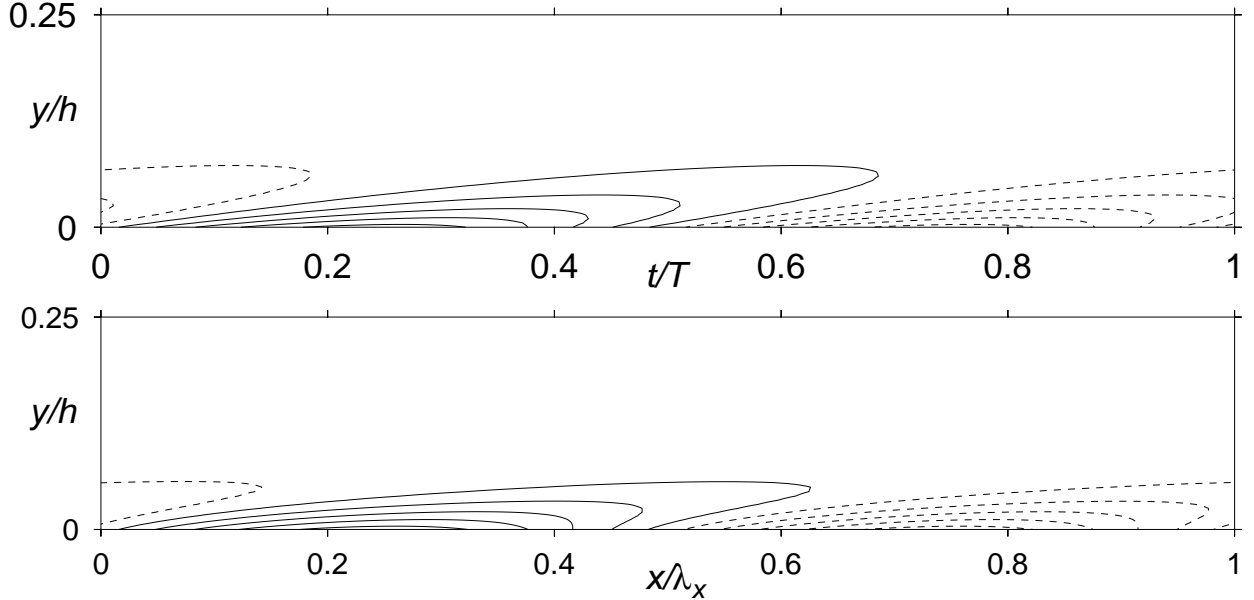}
\caption{Contour of the $w$ field for the laminar TSL and SSL. Top: temporal forcing $w(0,t)=\sin(\omega t)$. Bottom: spatial forcing $w(0,x)=\sin(\kappa x)$. Contours at $(-0.45,0.1,0.45)$, dashed lines indicate negative values.}
\label{fig:wcont}
\end{figure}

In both the TSL and SSL, the field of the spanwise velocity component $w$ is a function of two independent variables, namely $t,y$ for TSL and $x,y$ for the SSL. We start by observing in fig.\ref{fig:wcont} how closely these $w$ fields for TSL and SSL resemble each other. The abscissa in fig.\ref{fig:wcont} is $t/T$ for the TSL, whereas it is changed to $x / \lambda_x$ for the SSL. Both fields are computed with the DNS code, although the analytical expression of the former is available \cite{schlichting-gersten-2000}:
\begin{equation}
\label{eq:solution-tsl}
w(t,y) = C_t \Re \left[ \rme^{i \omega t} \rme^{-y / \delta_t} \right] ,
\end{equation}
where $C_t$ is a (real) normalization constant, $\omega = 2 \pi / T$ is the angular frequency of the oscillation and $\delta_t$ is the thickness of the TSL, defined as:
\begin{equation}
\label{eq:deltat}
 \delta_t = \left( \frac{ \nu }{ \omega } \right)^{1/2}.
\end{equation}

Figure \ref{fig:wcont}, where the parameters $T$ and $\lambda_x$ are in the range of interest for turbulent DR purposes, illustrates the wall-normal structure of the TSL and the SSL. Within this parameter range, both Stokes layers are thin compared to the channel half-width $h$. The convection velocity being approximately constant over such a small wall-normal extension explains why the two contours in figure \ref{fig:wcont} look very similar.

An analytical solution for the SSL can be arrived at under the small thickness approximation. After some analytical efforts, described in the Appendix, the $w(x,y)$ field of the laminar SSL can be shown to obey an expression, similar to Eq.(\ref{eq:solution-tsl}), that contains an Airy function instead of an exponential function:
\begin{equation}
 \label{eq:solution-ssl}
w(x,y)= C_x \Re \left[ \rme^{i \kappa x} 
\Ai \left( - \frac{i y}{\delta_x} \rme^{-i 4/3 \pi} \right)  \right] ,
\end{equation}
where $C_x$ is a (real) normalization constant, $\kappa = 2 \pi / \lambda_x$ is the forcing wavenumber, $\Ai$ is the Airy function of the first kind \cite{bender-orszag}, and $\delta_x$ is the thickness of the SSL, defined as:
\begin{equation}
\label{eq:deltax}
 \delta_x = \left( \frac{ \nu } { u_{y,0} \kappa } \right)^{1/3} .
\end{equation}

In this expression, $u_{y,0}$ represents the gradient of the streamwise mean velocity profile evaluated at the wall. The $w(x,y)$ laminar field computed by DNS and plotted in fig.\ref{fig:wcont} (bottom) is virtually indistinguishable from the same field as computed from the analytical solution (\ref{eq:solution-ssl}).

Comparing the two expressions (\ref{eq:solution-tsl}) and (\ref{eq:solution-ssl}), a qualitative difference between the SSL and the classical TSL can be observed: the former is not decoupled from the longitudinal mean velocity profile $u(y)$, but it depends on $u_{y,0}$ through the thickness $\delta_x$ given by (\ref{eq:deltax}). This remains without consequences in the laminar case, where the streamwise flow is not affected by the wall forcing, whereas it will become important in the turbulent case.

\subsection{Comparison with the mean spanwise turbulent flow}

\begin{figure}
\includegraphics[width=\columnwidth]{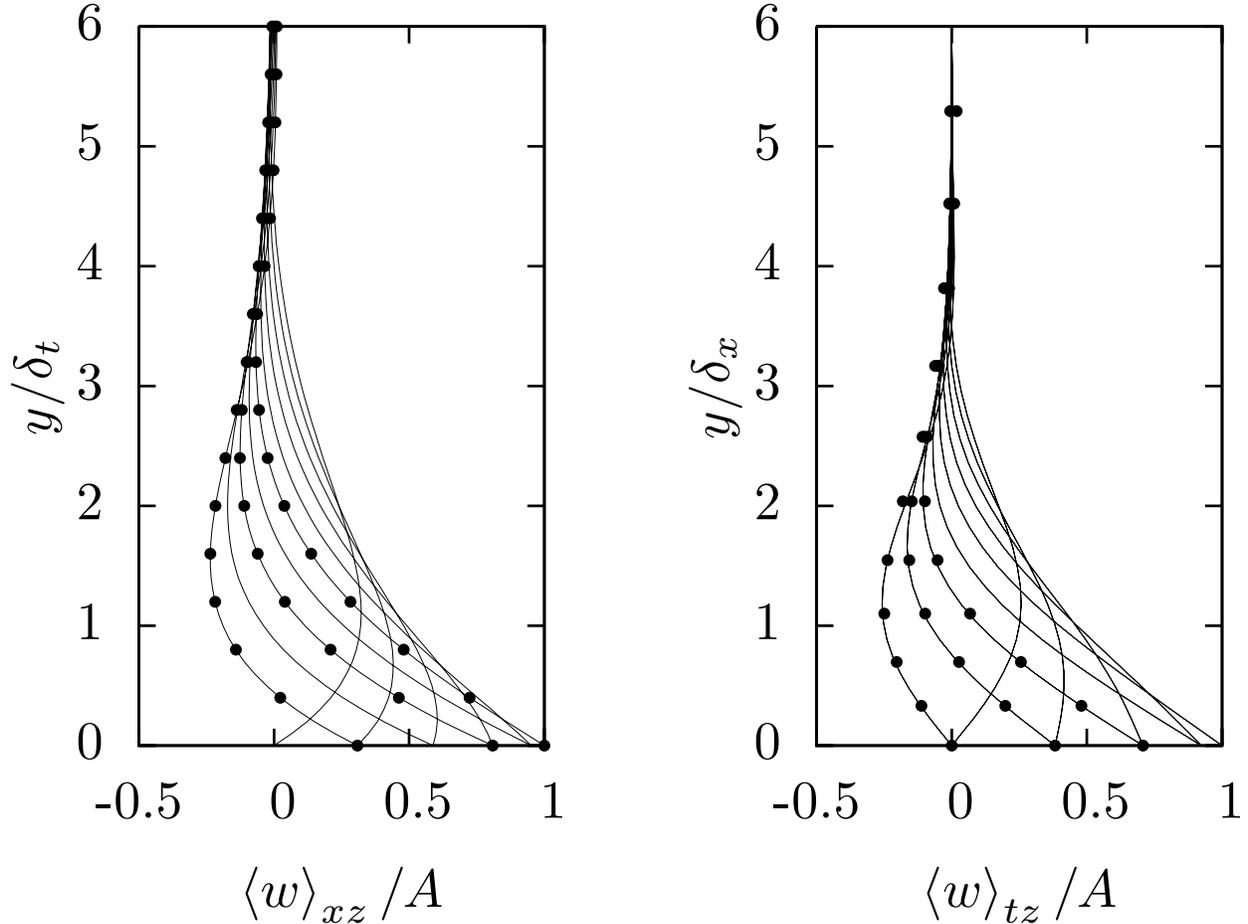}
\caption{Comparison between laminar and turbulent mean $w$ fields for temporal and spatial forcing. Left: analytical solution (lines) of the TSL and turbulent $\aver{w}_{xz}$ field (symbols) computed by DNS. Right: analytical solution (lines) of the SSL and turbulent $\aver{w}_{tz}$ field (symbols) computed by DNS. Profiles are taken at different (temporal and spatial) phases during the oscillation. The laminar solutions are linear in $A$, whereas the turbulent profiles are for $A^+=12$. Turbulent temporal case is at $Re_\tau=200$ and $T^+=125$. Turbulent spatial case is at $Re_\tau=200$ and $\lambda_x^+=1250$. To emphasize collapse with proper scaling, the SSL analytical solution is computed for a different wavelength, namely $\lambda_x^+=1900$, yielding a different $\delta_x$, so that the role of $y/\delta_x$ as similarity variable is highlighted.}
\label{fig:w-lamturb}
\end{figure}

It is well documented \cite{quadrio-ricco-2003} for the time-oscillating wall that the phase-averaged $\aver{w}_{xz}(y)$ profile is identical to the laminar solution expressed by Eq.(\ref{eq:solution-tsl}), except for the initial transient where the oscillation is started from rest. (The operator $\aver{\cdot}_{xz}$ indicates averaging along the homogeneous directions $x$ and $z$.) This is well illustrated by fig.\ref{fig:w-lamturb} (left), where the wall-normal distribution of $w$ in the TSL after Eq.(\ref{eq:solution-tsl}) is plotted at various phases during the cycle, and compared with the turbulent $\aver{w}_{xz}$ field. The agreement, which has been related by Ricco \& Quadrio \cite{ricco-quadrio-2008} to the vanishing $y,z$ component of the Reynolds stresses tensor, is, as expected, excellent.

The same result is shown in fig.\ref{fig:w-lamturb} (right) to hold true for the SSL case. Now, of course, the analytical solution (\ref{eq:solution-ssl}) has to be compared with the turbulent $\aver{w}_{tz}$ field. The role of $y/\delta_t$ as similarity variable is well known for the TSL. To emphasize that $y/\delta_x$ plays the same role for the SSL, the comparison between laminar and turbulent profiles is carried out for two cases with different $\kappa$ and thus, according to (\ref{eq:deltax}), with different $\delta_x$. The agreement between the two profiles is excellent. This confirms the analogy between TSL and SSL in terms of the connection between the laminar solution and the phase-averaged turbulent solution. This analogy can be used (see \ref{sec:results}) to increase our predictive capabilities  when the spatial forcing (\ref{eq:space}) is applied to a turbulent flow.

\section{Numerical method for DNS of turbulent flow} 
\label{sec:num}

We turn now to the turbulent case, that will be dealt with by using the Direct Numerical Simulation (DNS) technique. The computer code used to solve the incompressible Navier--Stokes equations is a parallel DNS solver, based on mixed discretization (Fourier expansion in the homogeneous directions, and compact fourth-order accurate explicit compact finite difference schemes in the wall-normal direction), recently developed by Luchini and Quadrio \cite{luchini-quadrio-2006}. The turbulent flow in an indefinite plane channel is considered, with gap half-width given by $h$. One reference simulation without forcing has been carried out at $Re_\tau=200$, where $Re_\tau$ is the friction Reynolds number, defined based on $h$ and the friction velocity $u_\tau$, that will be used as reference velocity throughout the paper unless otherwise noted. The size of the computational domain is $L_x = 6 \pi h$, $L_y= 2 h$ and $L_z = 3 \pi h$. The total averaging time is rather large, to allow for well converged time-averaged results, and amounts to approximately $10^4$ viscous time units. The employed number of modes / points is given by $N_x=320$, $N_y=160$ and $N_z = 320$, that yields a standard spatial resolution. For the reference simulation a value of $C_f=7.94 \cdot 10^{-3}$ (defined based on the bulk velocity) is obtained, in agreement with the correlation $C_f = 0.0336 Re_\tau^{-0.273}$ given by Pope \cite{pope-2000}.

Some 40 simulations with Eq.(\ref{eq:space}) used as boundary condition at the channel walls have been then carried out for different values of $\lambda_x$ and $A$. Owing to the periodic boundary conditions employed in the homogeneous directions, an integer number of forcing wavelength $\lambda_x$ must be contained in the domain length $L_x$. The systematic variation of $\lambda_x$ has thus required, mostly for the largest values of $\lambda_x$, slight adjustments of $L_x$. Compared to the reference value of $L_x=6 \pi h$, 3 cases had an actual value of $L_x = 7 \pi h$, and one was at $L_x = 5 \pi h$. Since such changes are of limited entity, the number of streamwise Fourier modes has been left unchanged, with the implied little change in streamwise spatial resolution for these few cases. Properly accounting for the initial transient, where the wall friction decreases from the unperturbed value to the reduced value given by the forcing, is enforced according to the procedure described by Quadrio \& Ricco \cite{quadrio-ricco-2004}, who reported an overall error in determining the friction coefficient below 1\%. We have also verified that the results are not affected by the chosen value of $L_x$. One simulation with $\lambda_x^+ = 1875$ and large DR has been repeated by doubling $L_x$ at $L_x = 12 \pi h$ (and of course doubling the number of streamwise Fourier modes). The value of $C_f$ measured in the simulation with the longer domain differs from the one computed by the simulation with standard domain length by much less than the error, mentioned above, related to the finite averaging time.

The simulations were run on a computing system available in dedicated mode at the Universit\`a di Salerno, taking advantage of its large computing throughput to run several cases at a time. The system possesses 64 computing nodes, each of which is equipped with two dual-core AMD Opteron CPUs. The single computational case took about 10 days of wall-clock time when run in parallel by using 8 nodes. Up to 8 cases can be run simultaneously, so that the wall-clock time for the entire study was about 7 weeks.

\section{Turbulent flow}
\label{sec:results}

The effectiveness of the steady spatial forcing (\ref{eq:space}) in reducing the frictional drag is assessed by examining the value of the skin-friction coefficient obtained in several different simulations, in which the amplitude $A$ and the wavelength $\lambda_x$ of the wall forcing are systematically varied. The measured coefficients are then compared to the value of the reference (unforced) flow. The savings in driving power will then be compared to the energetic cost of the wall velocity distribution, in order to asses the net power saving made possible by the spatial forcing.

\subsection{Pumping power saved}
\label{sec:Psav}
Our simulations are performed at a fixed flow rate, so that a decrease of the frictional drag translates into a proportional decrease of the mean streamwise pressure gradient and of the power required to drive the flow. The power to drive the flow against viscous resistance is defined as:
\[
\sP_{dr} = \frac{U_bL_xL_z}{t_f-t_i}
\int_{t_i}^{t_f} \left( \tau_{x,\ell} + \tau_{x,u} \right) \ud t,
\]
where $\tau_{x,\ell}$ and $\tau_{x,u}$ are the space-averaged value of the streamwise component of the wall shear stress, evaluated at the lower and upper wall respectively, and $(t_f - t_i)$ is the time interval over which the time averaged is carried out, after discarding initial transients.

The percentage saved power $P_{sav}$ is expressed as percentage of the power $\sP_{dr,0}$ required to drive the flow in the reference case, and is defined as follows:
\[
P_{sav} = 100 \frac{\sP_{dr,0} - \sP_{dr}}{\sP_{dr,0}}.
\]

The quantity $P_{sav}$ exactly corresponds to the percentage of friction drag reduction, and is expected to be a function of $A$ and $\lambda_x$.

\begin{figure}
\centering
\includegraphics[width=\columnwidth]{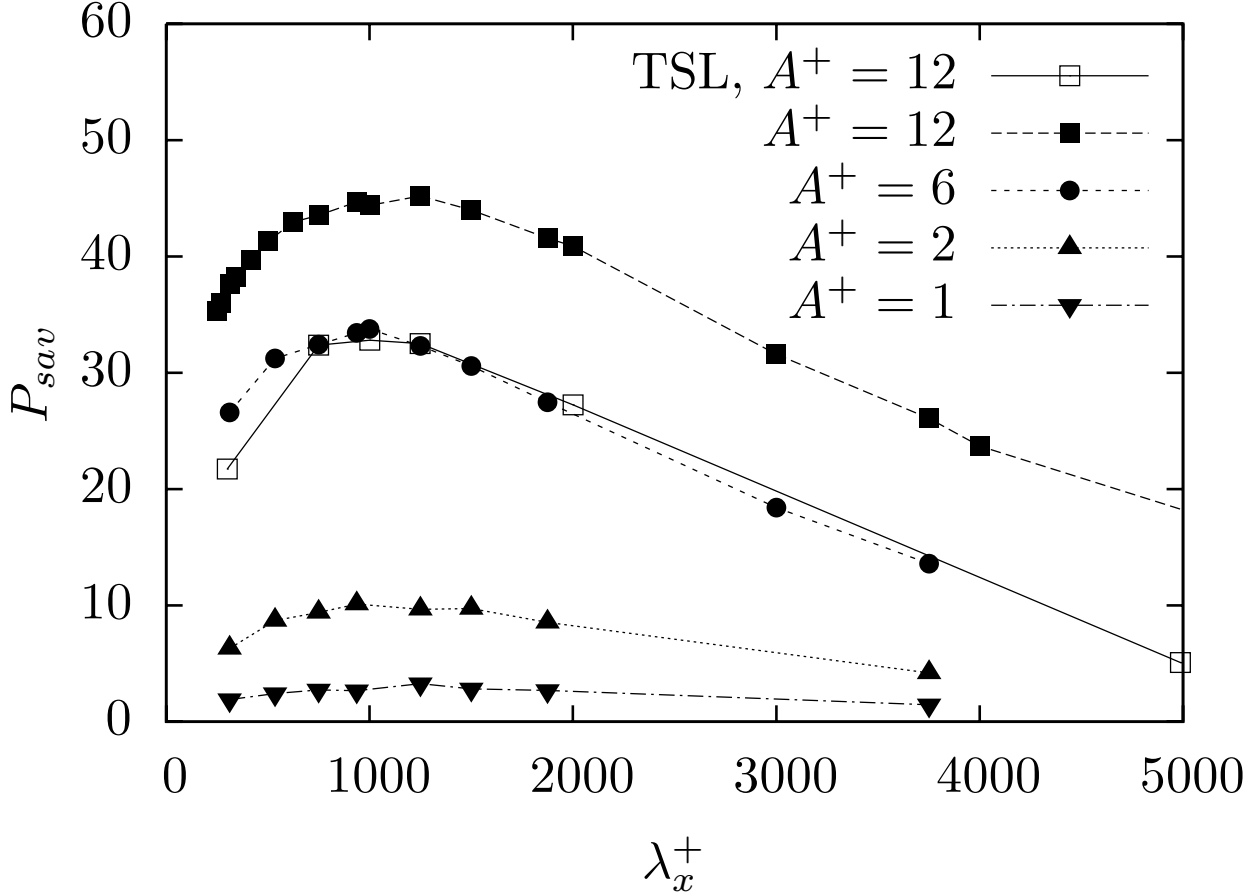}
\caption{Percentage drag reduction $P_{sav}$ as a function of the forcing wavelength. Empty symbols are temporal cases from Quadrio \& Ricco \cite{quadrio-ricco-2004}: for them $\lambda_x^+ = \sU_w^+  T^+$ (see text).}
\label{fig:Psav-lambda}
\end{figure}
In fig.\ref{fig:Psav-lambda} $P_{sav}$ is plotted first as a function of $\lambda_x^+$, for different values of the amplitude $A^+$. Available data \cite{quadrio-ricco-2004} obtained with temporal forcing are also plotted, with the oscillation period translated into an oscillation wavelength through the near-wall value of the convection velocity, namely $\sU_w^+=10$.

The capital information drawn from this plot is the validation of our working hypothesis: the present steady forcing indeed parallels the unsteady oscillating-wall technique when oscillation period and forcing wavelength are related through $\sU_w$. In analogy to the oscillating wall, that yields the maximum DR at a well-defined oscillation period $T^+$, namely $T^+_{opt} = 100 - 125$, the steady forcing yields the maximum DR at a well-defined wavelength $\lambda_x^+$, namely $\lambda^+_{x,opt} = 1000 - 1250$. It is striking how well the prediction $\lambda^+_{x,opt}= \sU_w^+ T^+_{opt}$ is confirmed by our simulations. DR is observed over a very wide range of wavelenghts, $200 < \lambda_x^+ < 8000$, analogously to the oscillating wall.

The effects of the SSL on the turbulent flow are, however, only qualitatively similar to those of the TSL. Quantitative differences do exist, and in particular at a given wavelength the spatial forcing is observed to attain larger values of DR when compared to the temporal forcing at the same amplitude and equivalent period.
\begin{figure}
\centering
\includegraphics[width=\columnwidth]{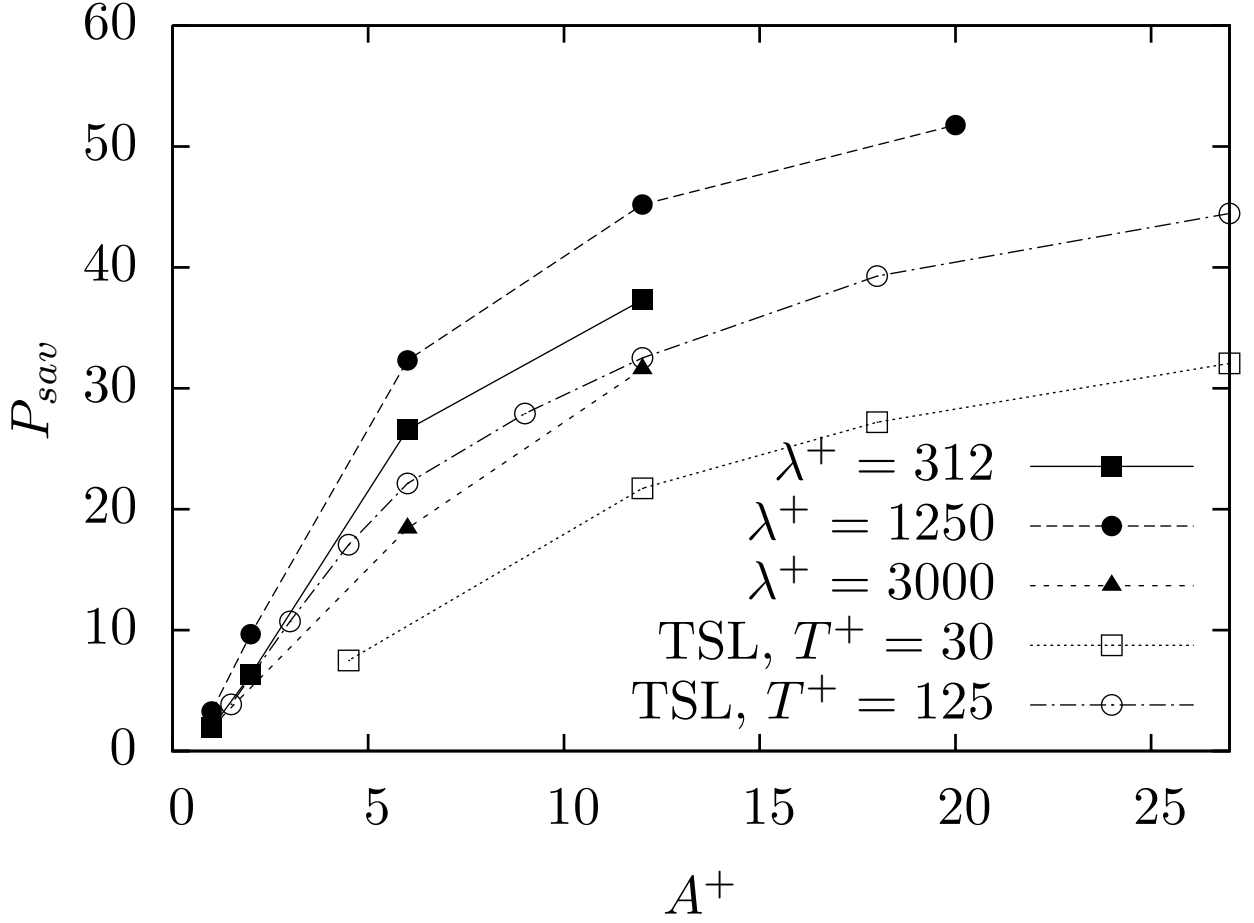}
\caption{Percentage drag reduction $P_{sav}$ as a function of forcing amplitude. As in fig.\ref{fig:Psav-lambda}, empty symbols are temporal cases from Quadrio \& Ricco \cite{quadrio-ricco-2004}: for comparison recall that $T^+ = \lambda_x^+ / \sU_w^+$ (see text).}
\label{fig:Psav-A}
\end{figure}
Fig.\ref{fig:Psav-A} shows $P_{sav}$ as a function of the amplitude $A^+$, for different values of $\lambda_x^+$. The plot contains data from spatial as well as from temporal forcing cases. The dependence of DR on the forcing amplitude is qualitatively very similar for TSL and SSL: DR monotonically grows with $A$, but the increase saturates to an apparently asymptotic behavior. Again, the space-dependent forcing appears to be more effective in reducing turbulent friction: for a given amplitude the spatial forcing yields a $20-30\%$ larger DR than the temporal forcing. A maximum DR of about $52\%$ is observed at $\lambda_x^+=1250$ with a forcing amplitude of $A^+=20$.

\subsection{Power expended at the wall}
\label{sec:Preq}

In addition to $\sP_{dr}$, the forced channel flow has an additional power input $\sP_{req}$, that is required to enforce the wall motion against the spanwise shear stress. $\sP_{req}$ is defined as:
\begin{equation}
\label{eq:preq}
\sP_{req} = \frac{L_z}{t_f-t_i} \int_{t_i}^{t_f} \int_0^{L_x}
W \left( \tau_{z,\ell} + \tau_{z,u} \right) \ud x \ud t ,
\end{equation}
where $\tau_z$ is the space-averaged value of the spanwise component of the wall shear stress, and $W$ is the spanwise velocity of the walls.

\begin{figure}
\centering
\includegraphics[width=\columnwidth]{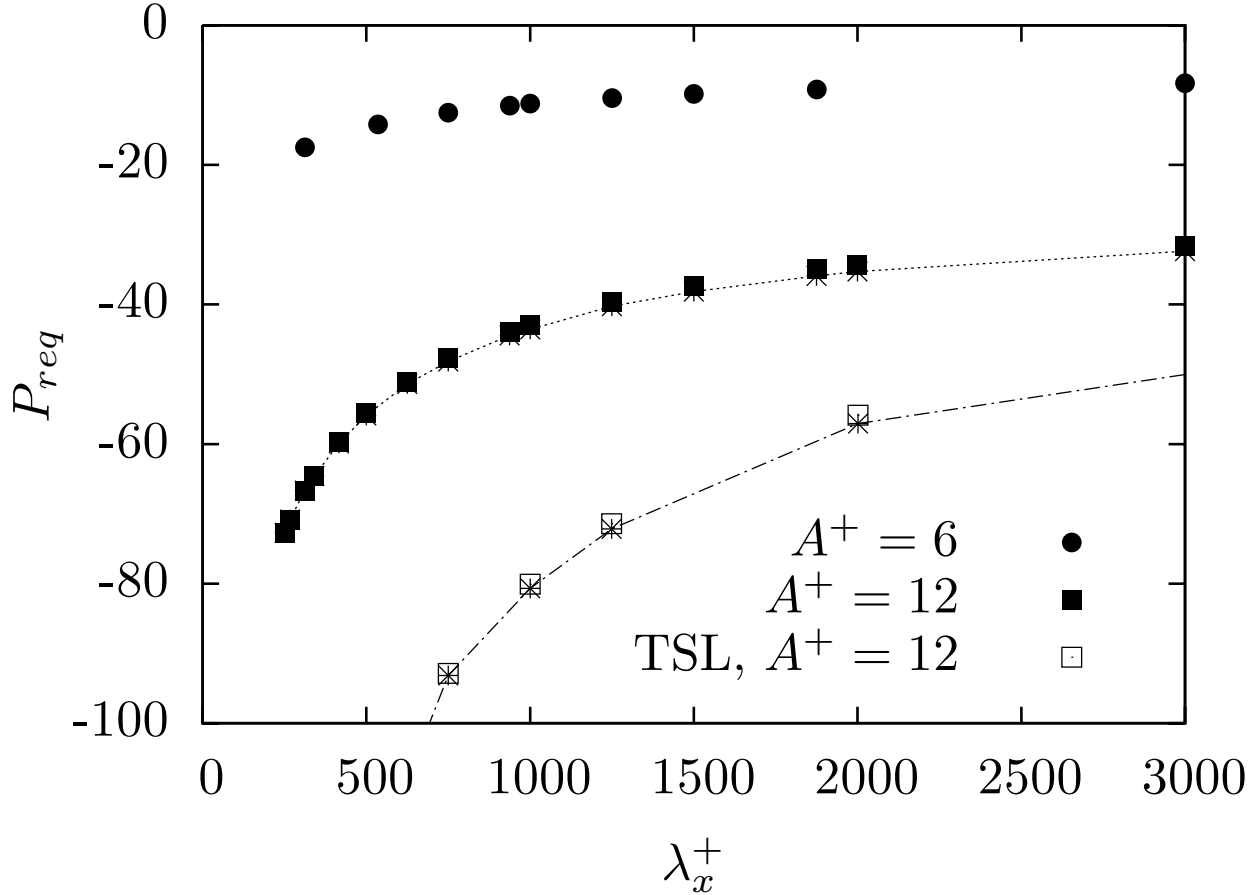}
\caption{Percentage required power $P_{req}$ as a function of the forcing wavelength. As in fig.\ref{fig:Psav-lambda}, empty symbols are temporal cases from Quadrio \& Ricco \cite{quadrio-ricco-2004}. The line-connected asterisks represent $P_{req}$ obtained from the laminar solution with $A^+=12$, from formula (\ref{eq:solution-tsl}) for the TSL and from formula (\ref{eq:solution-ssl}) for the SSL.
\label{fig:Preq-lambda}}
\end{figure}

The required percentage power $P_{req}$ is defined in terms of the friction power $\sP_{dr,0}$ of the reference flow as $P_{req} = \sP_{req} / \sP_{dr,0}$. $P_{req}$ is presented in fig.\ref{fig:Preq-lambda} as a function of the forcing wavelength. Again, spatial as well as temporal forcing data are included. Of course $P_{req}$ assumes negative values, i.e. work has to be done against the fluid viscosity. Comparing the two forcing methods, one can easily appreciate how the spatial forcing presents an energetic cost that is smaller than the cost of temporal forcing, by approximately a factor of 2. 

The line-connected points in fig.\ref{fig:Preq-lambda} represent the values $P_{req}$ that can be computed from laminar theory. TSL and SSL are considered at $A^+=12$, for which formulas (\ref{eq:solution-tsl}) and (\ref{eq:solution-ssl}) respectively are analytically integrated. The good agreement was expected, since it was already observed (see fig.\ref{fig:w-lamturb}) that laminar and turbulent mean profiles of spanwise velocity are coincident.  

\begin{figure}
\centering
\includegraphics[width=\columnwidth]{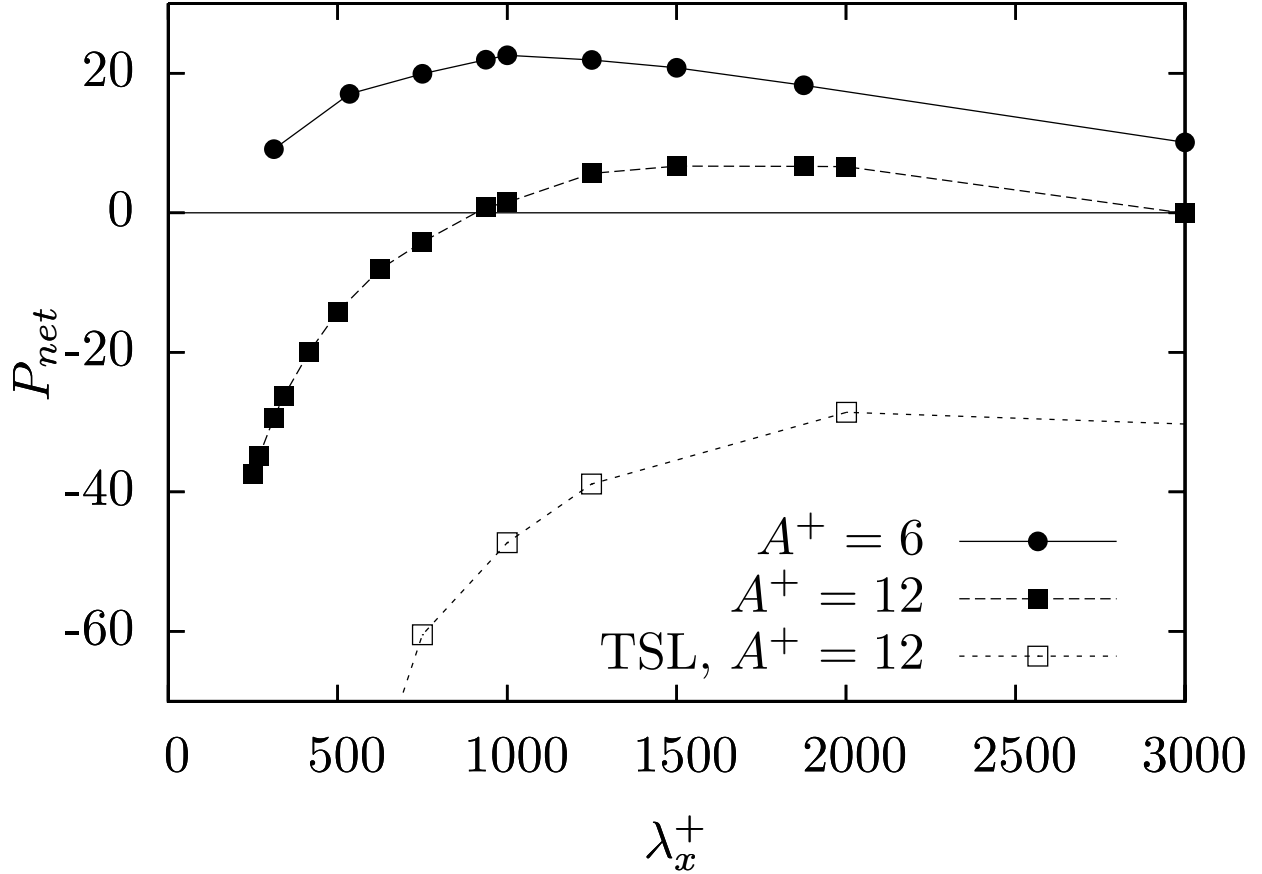}
\caption{Percentage net power saving $P_{net}$ as a function of the forcing wavelength. 
\label{fig:Pnet-lambda}}
\end{figure}

Lastly, a net percentage power saving $P_{net}$ is easily defined by comparing $P_{req}$ and $P_{sav}$, as follows:
\[
P_{net}= P_{sav} + P_{req}.
\]

The net power saving $P_{net}$ is plotted in fig.\ref{fig:Pnet-lambda}. Since it has already been observed how SSL yields higher DR at lower energetic cost, of course here a significant difference in net gain is expected when comparing SSL and TSL. $P_{net}$ shows indeed an interesting maximum value of about $23\%$ at $A^+=6$, while remaining positive over a wide range of wavelengths. Moreover, positive $P_{net}$ are found for rather large amplitudes: at $A^+=12$ a net gain of about 5\% can still be observed, whereas the TSL at the same amplitude presents a net loss of about 30\%. It is worth noting that the search for the maximum of $P_{net}$ cannot be considered exhaustive, and thus the presently observed maximum value of 23\% at $A^+=6$ should be regarded as a starting point for a refined search.

\subsection{Flow statistics}
\label{sec:flowviz}

\begin{figure}
\centering
\includegraphics[width=\columnwidth]{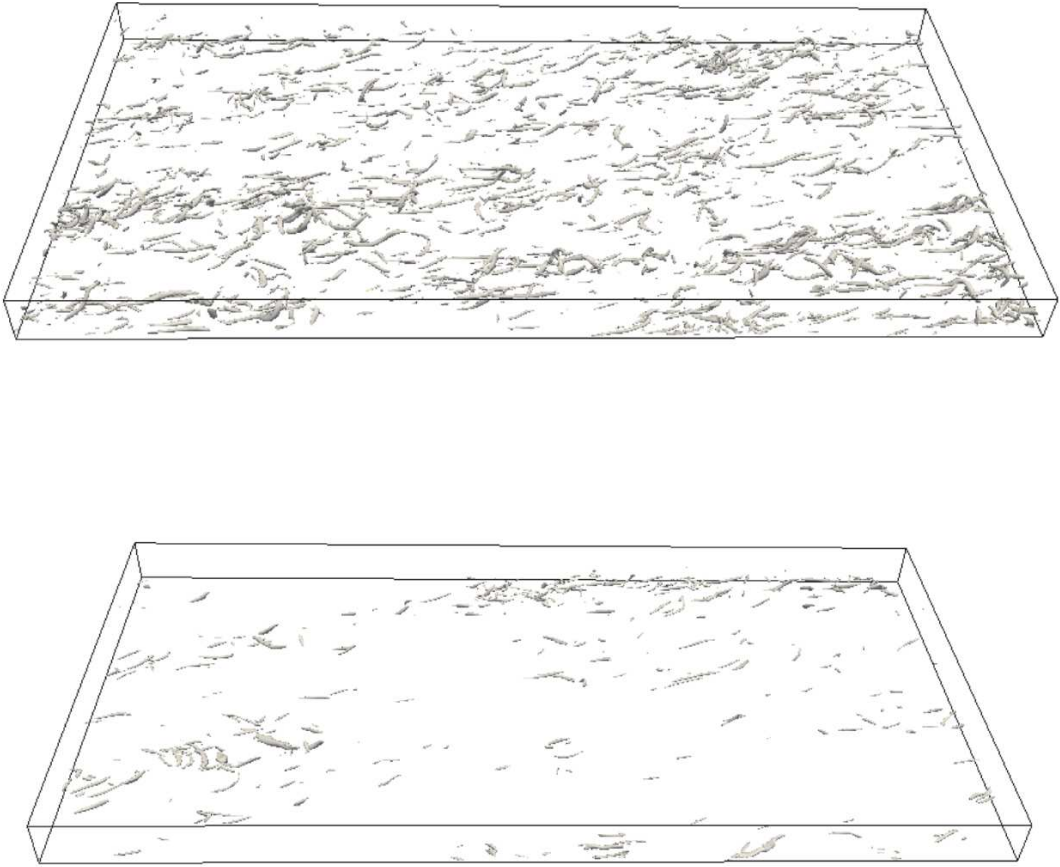}
\caption{Isosurfaces for the quantity $\lambda_2^+=-0.03$.  Flow is from left to right. Top: reference flow; bottom: spatial forcing with $\lambda_x^+=1250$ and $A^+=12$.}
\label{fig:ucont}
\end{figure}

Observing the main statistics of the flow and even a few instantaneous snapshots may help understanding how SSL affects turbulence. (Additional statistical quantities are presented elsewhere \cite{quadrio-viotti-luchini-2007}). In fig.\ref{fig:ucont} isosurfaces are visualized for the $\lambda_2$ quantity introduced by Jeong \& Hussain \cite{jeong-hussain-1995} and since then often used to identify turbulent vortical structures. The level is set at $\lambda_2^+=-0.03$. The top plot is for the reference flow, and the bottom plot is for the same flow subject to the effects of the spatial forcing, with $\lambda^+ = 1250$ and $A^+ = 12$. This is one of the cases with the highest drag reduction, i.e. 45\%. The SSL clearly appears to modify the near-wall turbulence and its structures.

\begin{figure}
\centering
\includegraphics[width=\columnwidth]{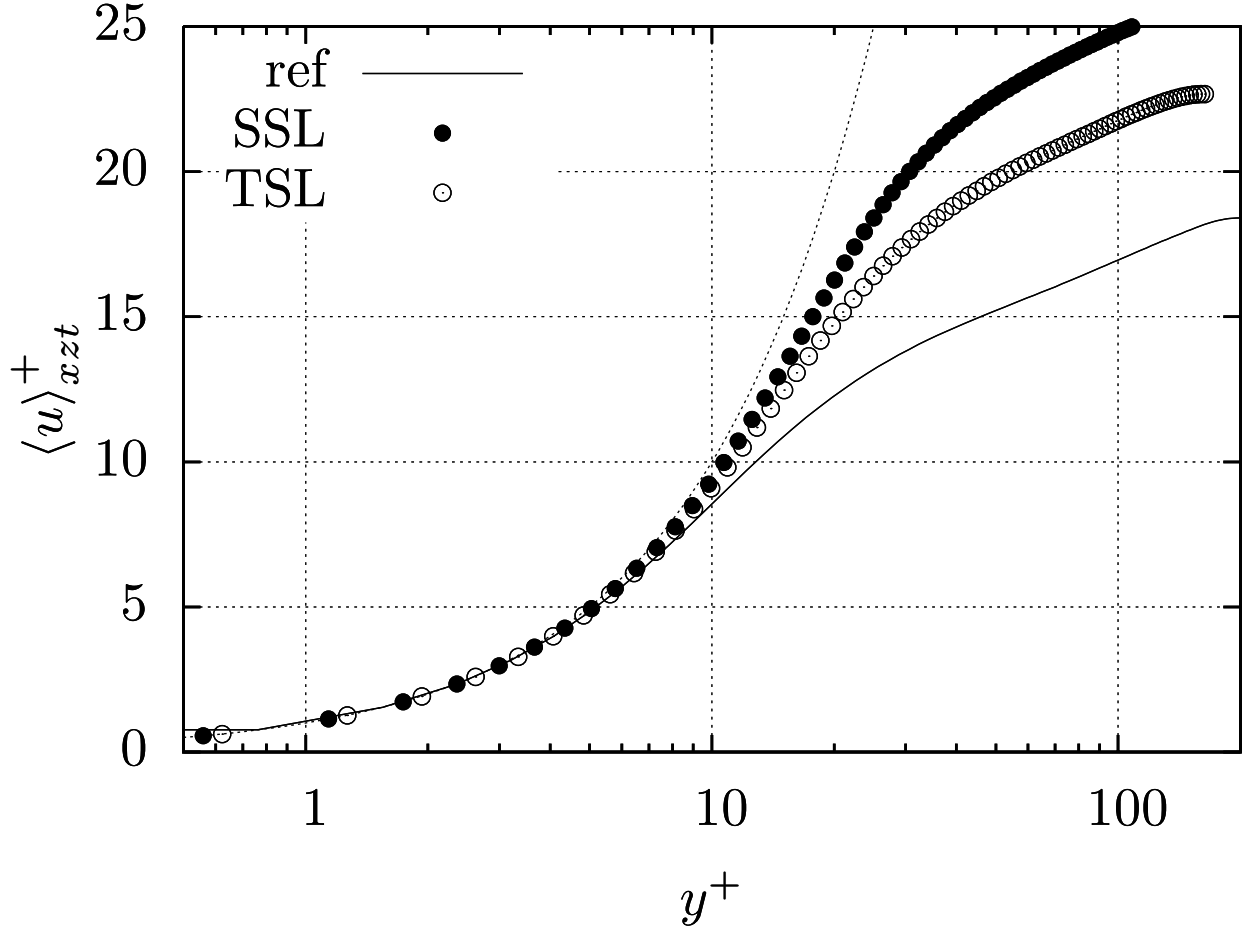}
\caption{Streamwise mean velocity profile in law-of-the-wall form. Dotted line is the linear velocity profile $\aver{u}^+_{xzt} = y^+$.}
\label{fig:uav}
\end{figure}

Similar considerations can be drawn from single realizations of the flow as well from its statistical description. Fig.\ref{fig:uav} reports the mean velocity profile in the law-of-the-wall form. The modification to the profile for the TSL and SSL cases are analogous when compared to the reference flow, but the effects are larger for the SSL. The drag reduction manifests itself through the thickening of the viscous sublayer, that results in the upward shift of the logarithmic portion of the velocity profile, as previously documented for other DR techniques, for example riblets \cite{choi-1989}.

\begin{figure}
\centering
\includegraphics[width=\columnwidth]{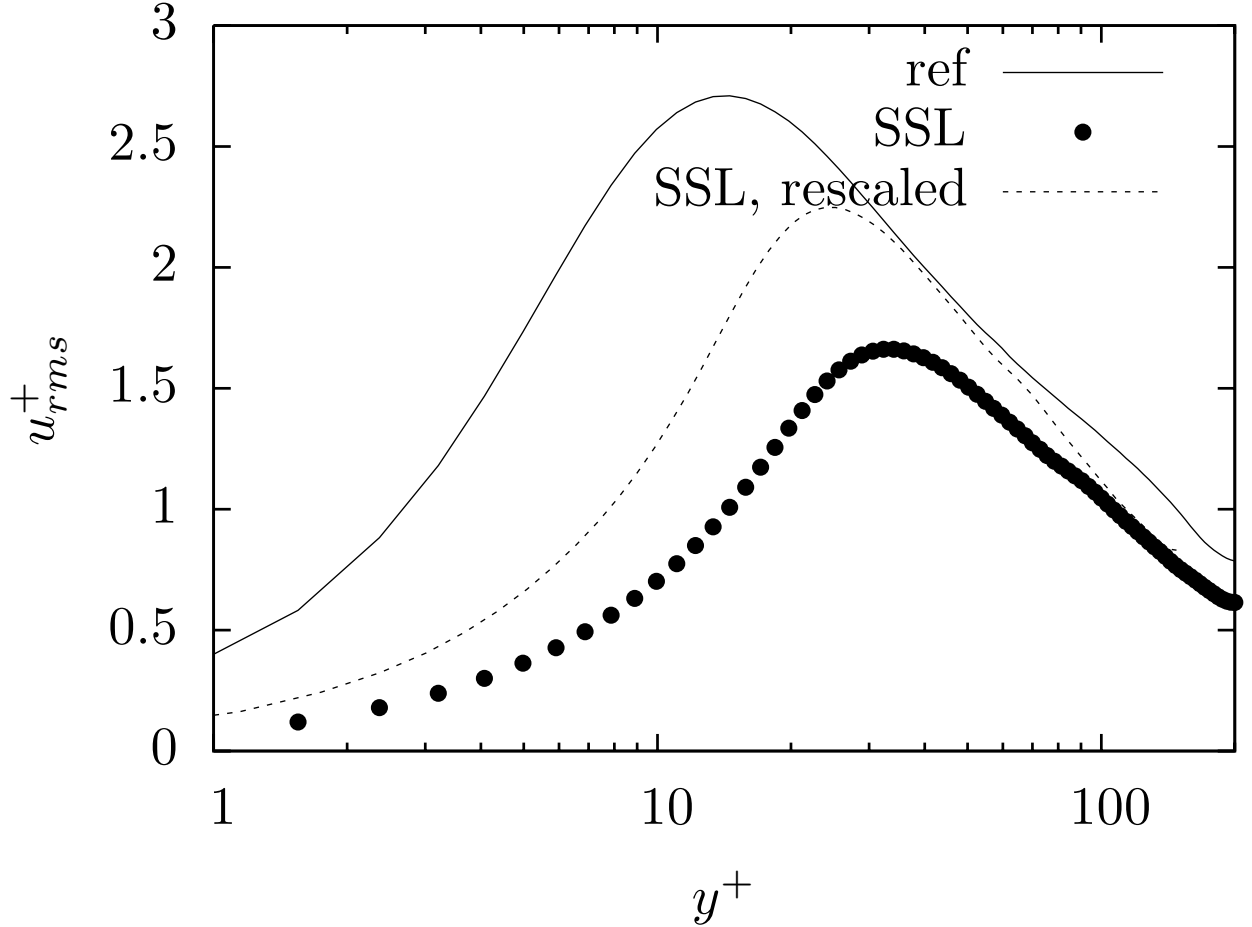}
\caption{Root-mean-square value of the fluctuations for the streamwise velocity component.}
\label{fig:urms}
\end{figure}

Another relevant statistical quantity that is modified by the action of the SSL is the turbulence intensity. Fig.\ref{fig:urms} presents the r.m.s. value of the fluctuations of the streamwise velocity component. To examine the effect of the lower Reynolds number implied by the strong drag reduction in a constant-flow-rate simulation, this plot is still scaled using inner variables, but both the friction velocity of the reference flow and the actual friction velocity of the drag-modified flow are used as velocity scale. The strong reduction of fluctuations is deemphasized with the latter form of scaling, and what remains is the structural effect of the SSL after subtracting the effect due to a smaller $Re_\tau$. The curves however remain significantly different; the main effect is a decrease of the fluctuation intensity, together with a displacement of the position of the maximum intensity further form the wall. Similar effects have been already documented for other DR methods. In particular the same observation has been put forward \cite{quadrio-ricco-2004} for the oscillating-wall. It may be useful to remind the reader that effective DR methods exist, for example the active opposition control \cite{choi-moin-kim-1994}, where $u_{rms}^+$ is unaffected when properly scaled; on the other hand, as discussed for example by Jim\'enez \cite{jimenez-2004}, experimental evidence exists that wall roughness may reduce the near-wall peak of turbulence intensities while increasing drag.

\section{Discussion and conclusions}
\label{sec:conclusions}

This paper has studied a new form of boundary forcing for wall-bounded turbulent flows, that consists in imposing at the wall a steady distribution of spanwise velocity, modulated in the streamwise direction. In this study only sinusoidal modulation has been considered. Main motivation was to find a steady counterpart to the oscillating-wall technique. The link between the two kinds of forcing is the convection velocity of the turbulence fluctuations, that takes a well-defined non-zero value $\sU_w$ at the wall and is thus capable of transforming a time scale into a length scale and viceversa.

Thanks to a number of direct numerical simulations, the behavior of this new forcing in the parameter space has been determined, and DR up to $52\%$ has been observed for $A^+=20$ and $\lambda_x^+=1250$. For all amplitudes, the forcing wavelength that yields the maximum DR has been found to correspond to the optimal period of the oscillating wall converted in length through $\sU_w$, thus confirming the validity of the analogy between temporal and spatial forcing.

This analogy has been extended further by studying the laminar case: this was known to be relevant to the oscillating-wall technique, since in the turbulent case the spanwise profile after space-time averaging is identical to the laminar solution. The laminar solution, that can be written in terms of Airy functions, has been determined for the spatial case too, and it has been additionally verified that the turbulent spanwise flow when phase-averaged is identical to the laminar solution. This property can be leveraged to predict the power required for the control. We hope that a predictive quantity, similar to the parameter discussed by Quadrio \& Ricco \cite{quadrio-ricco-2004,ricco-quadrio-2008} and capable of describing the DR effects of the forcing, could be envisaged on the basis of this analytical solution.

Together with qualitative analogies, there are quantitative differences between temporal and spatial forcing. The spatial forcing is more efficient in terms of DR, from the point of view of both absolute DR and net power saving. In particular, a net power saving as high as $23\%$ has been computed at $A^+=6$, with one unit of forcing power translating into 3 units of saved power. This is more than 3 times the largest net saving documented for the oscillating wall, and also significantly larger than the benefit obtained with passive devices like riblets, that are reportedly capable of a saving up to 8-10\% in laboratory conditions \cite{bechert-etal-1997}.

The present form of spatial forcing is certainly realizable in principle, and an experimental setup is indeed under construction that will help shedding light on effects like the dependence of DR on $Re$. At the same time, we do not consider the present forcing, though steady, directly suitable as yet for practical applications. However, the successful design of a steady control law is one important step towards the realization of a passive drag-reducing device. In this framework, the sensitivity of the turbulence near-wall cycle to a well-defined streamwise lengthscale is a fundamental result, that paves the way to the search for an efficient drag-reducing type of roughness. In this context, we have recently applied for a patent \cite{quadrio-luchini-2008} concerning a design method of roughness distributions yielding improved benefits over conventional riblets. Such distributions may contain riblets with sinusoidally ondulated crests with wavelength $\lambda_{opt}$. A later paper by Peet {\em et al.} \cite{peet-sagaut-charron-2008} has confirmed by numerical simulations the potential success of similar sinusoidal riblets, reporting a 50\% improvement over conventional riblets. Since in \cite{peet-sagaut-charron-2008} the wavelength was in the optimum range, but only a single amplitude was tested, we are confident that even larger benefits can be attained by such a purely passive technique.

\section{Acknowledgments}
CV has been supported by Italian Ministry of University and Research through the grant PRIN 2005 on {\em Large scale structures in wall turbulence}. We acknowledge interesting discussions with Dr P.Ricco. Part of this work has been presented by CV in June 2007 at the XI European Turbulence Conference, Porto (P).

\section{Appendix}
\label{sec:analytical}
In this Appendix an analytical, approximate solution of the Navier--Stokes equations for the laminar flow between indefinite plane walls is described, where the non-homogeneous boundary condition
\begin{equation}
\label{eq:bc-z}
w(x,0,z,t) = A \cos (\kappa x)
\end{equation}
is imposed to the spanwise velocity component. The boundary forcing creates a layer of alternate spanwise motion, which develops close to the wall and resembles the temporal Stokes layer. We have referred to it in this paper as the Spatial Stokes Layer (SSL), in comparison to the conventional Temporal Stokes Layer (TSL). An analytical solution will be derived now for the velocity profile of the SSL, under the assumption that the wall-normal length scale characteristic of the layer is small in comparison to the channel width.

The solution is steady, and thanks to the spanwise invariance of the differential system, including its boundary conditions, all the $z$ derivatives can be dropped from the momentum equations, which then read:
\begin{subequations}
\begin{equation}
\label{eq:N-S4}
u\pd{u}{x}+v\pd{u}{y} = -\frac{1}{\rho}\pd{p}{x} + \nu \left( \ppd{u}{x}+\ppd{u}{y} \right),
\end{equation}
\begin{equation}
\label{eq:N-S5}
u\pd{v}{x}+v\pd{v}{y} = -\frac{1}{\rho}\pd{p}{y} + \nu \left( \ppd{v}{x}+\ppd{v}{y} \right),
\end{equation}
\begin{equation}
\label{eq:N-S6}
u\pd{w}{x}+v\pd{w}{y} = \nu \left( \ppd{w}{x}+\ppd{w}{y} \right).
\end{equation}
\end{subequations}

This highlights that Eqs.(\ref{eq:N-S4}) and (\ref{eq:N-S5}) decouple from  Eq.(\ref{eq:N-S6}) to form an independent two-dimensional problem, unaffected by the inhomogeneous boundary condition (\ref{eq:bc-z}). Its solution is thus the classical laminar Poiseuille solution, that gives a parabolic longitudinal velocity profile, and predicts a wall-normal velocity $v \equiv 0$ everywhere. Thus Eq.(\ref{eq:N-S6}) can be further simplified as:
\begin{equation}
\label{eq:prob1}
u\pd{w}{x}=\nu\ppd{w}{x}+\nu\ppd{w}{y} ,
\end{equation}

In (\ref{eq:prob1}) $u=u(y)$ is the (known) parabolic Poiseuille profile. The PDE (\ref{eq:prob1}) is linear, and thus in the following we will consider $A=1$ without loss of generality.

At this point a boundary-layer approximation is introduced. We suppose the SSL to be confined in a thin region close to the wall, and to vanish at a distance definitely smaller than $h$. If $\delta_x$ indicates the characteristic thickness of the SSL, we are requiring that $\delta_x \ll h$. Even before giving a precise definition of $\delta_x$, we have seen in fig.\ref{fig:wcont} that this requirement is satisfied when the flow parameters are set within the range of interest (which is where the spatial forcing achieves a substantial DR in the turbulent regime).

For small $\delta_x / h$, $u(y)$ in Eq.(\ref{eq:prob1}) can be replaced with the first term in its Taylor expansion:
\[
u(y) \approx u_{y,0} y .
\]

If $\lambda_x$ is comparable or larger than $h$, the boundary-layer approximation implies also that $\partial^2 w / \partial x^2$ in Eq.(\ref{eq:prob1}) is negligible compared to $\partial^2 w / \partial y^2$. We are thus left with the well-posed problem:
\begin{equation}
\label{eq:prob2}
u_{y,0} y \pd{w}{x}=\nu\ppd{w}{y},
\end{equation}
with boundary conditions:
\begin{eqnarray} \label{eq:BC2}
w(x,0) & = & \cos (\kappa x) \\ \nonumber
\lim_{y\rightarrow\infty}w(x,y) & = & 0 \nonumber
\end{eqnarray}
to be solved in the domain $y \in (0,+\infty)$. 

Its general solution $w(x,y)$ has the form:
\begin{equation}
\label{eq:sobst}
 w(x,y) = \Re \left[  \rme^{i \kappa x} F(y) \right] ,
\end{equation}
where the function $F$ is complex valued, $F(y): \mathbf{R} \rightarrow \mathbf{C}$.

By substituting the functional form (\ref{eq:sobst}) into Eq.(\ref{eq:prob2}) an ordinary differential equation for the unknown function $F$ is obtained:
\begin{equation}
\label{eq:prob3}
i \nu^{-1} \kappa u_{y,0} y F(y) = \frac{\ud^2 F(y)}{\ud y^2} .
\end{equation}

Its boundary conditions follow directly from (\ref{eq:BC2}):
\begin{equation}
\label{eq:BC3}
 \Re [F(0)]=1, \quad \Im [F(0)]=0, \quad
\lim_{y \rightarrow \infty}F(y) = 0 .
\end{equation}

To simplify notation, the factors $\nu^{-1} \kappa u_{y,0}$ in the l.h.s. of Eq.(\ref{eq:prob3}) are written in terms of a single parameter $\delta_x$, which has dimensions of a length:
\begin{equation}
\nu^{-1} \kappa u_{y,0} = \delta_x^{-3} ,
\end{equation}

Eq.(\ref{eq:prob3}) then becomes:
\begin{equation} \label{eq:prob4}
i \delta_x^{-3} y F = \frac{\ud^2 F}{\ud y^2}.
\end{equation}

Introducing the change of variable $y= i \delta_x \ty$, and redefining the unknown function as $F(i \delta_x \ty) = \tilde{F}(\ty)$ turns Eq.(\ref{eq:prob4}) into the following Airy equation:
\begin{equation}
\label{eq:airy}
 \ty \tilde{F} (\ty) = \frac{\ud^2 \tilde{F}}{\ud \ty^2} .
\end{equation}

Infinite solutions of an Airy equation exist for $\ty$ spanning the whole complex plane, when derivatives are considered in the sense of analytic functions. These solutions are linear combinations of the two special functions $\Ai(\ty)$ and $\Bi(\ty)$
\begin{equation}
\label{eq:combination}
 \tilde{F}(\ty)= \alpha \Ai(\ty) + \beta \Bi(\ty) ,
\end{equation} 
which are called Airy functions of the first and second kind respectively. 
The general solution (\ref{eq:combination}) has an alternate representation \cite{bender-orszag}, which turns out to be useful in our case:
\begin{equation}
 \tilde{F}(\ty)= \gamma \Ai(\omega \ty) + \theta \Ai(\omega^2 \ty) ,
\end{equation}
where $\omega=\rme^{-i2/3\pi}$. It can be shown \cite{bender-orszag} that among $\Ai(\ty)$, $\Bi(\ty)$, $\Ai(\omega\ty)$ and $\Ai(\omega^2 \ty)$,  the only base function satisfying conditions (\ref{eq:BC3}) (up to a normalization factor) is $\Ai(\omega^2 \ty)$. As a consequence, the solution that satisfies the required boundary conditions is:
\begin{equation}
 \tilde{F}(\ty)= \frac{\Ai(\omega^2 \ty)}{\Ai(0)}.
\end{equation}

By substituting this solution into (\ref{eq:sobst}), the expression for the unknown function $w(x,y)$ is eventually derived:
\begin{equation}
w(x,y)= C_x \Re \left[ \rme^{i \kappa x} 
\Ai \left( - \frac{i y}{\delta_x} \rme^{-i4\pi/3} \right) \right]
\end{equation}
where $C_x = \Ai(0)^{-1}$.

\bibliographystyle{unsrt}

\end{document}